\documentclass[graybox]{svmult}

\usepackage{mathptmx}       % selects Times Roman as basic font
\usepackage{helvet}         % selects Helvetica as sans-serif font
\usepackage{courier}        % selects Courier as typewriter font
\usepackage{type1cm}        % activate if the above 3 fonts are
\usepackage{makeidx}         % allows index generation
\usepackage{graphicx}        % standard LaTeX graphics tool
\usepackage{multicol}        % used for the two-column index
\usepackage[bottom]{footmisc}% places footnotes at page bottom

\makeindex             % used for the subject index
\begin{document}

\title*{Role of multiple subband renormalization in the electronic transport of 
correlated oxide superlattices}
\titlerunning{Subband renormalization in correlated oxide superlattices}

\author{Andreas R\"uegg and Manfred Sigrist}


\institute{Andreas R\"uegg \and Manfred Sigrist\at Theoretische Physik, ETH Z\"urich, 8093 Z\"urich, Switzerland. \email{rueegga@phys.ethz.ch}}

\maketitle

\abstract*{Metallic behavior of band-insulator/ Mott-insulator interfaces was observed
  in artificial perovskite superlattices such as in nanoscale
  SrTiO3/LaTiO3 multilayers. Applying a semiclassical perspective to the parallel electronic transport we identify two major
  ingredients relevant for such systems: i) the quantum confinement of the
  conduction electrons (superlattice modulation) leads to a complex, quasi-two dimensional subband
  structure with both hole- and electron-like Fermi surfaces. ii) strong
  electron-electron interaction requires a substantial renormalization of
  the quasi-particle dispersion. We characterize this renormalization by two sets of parameters, namely, the quasi-particle weight and the induced particle-hole asymmetry of each partially filled subband. 
In our study, the quasi-particle dispersion is calculated self-consistently as function of microscopic parameters using the slave-boson mean-field approximation introduced by Kotliar and Ruckenstein. 
We discuss the consequences of strong local correlations on the normal-state free-carrier response in the optical conductivity and on the thermoelectric effects.}

\abstract{Metallic behavior of band-insulator/ Mott-insulator interfaces was observed
  in artificial perovskite superlattices such as in nanoscale
  SrTiO$_3$/LaTiO$_3$ multilayers. Applying a semiclassical perspective to the parallel electronic transport we identify two major
  ingredients relevant for such systems: i) the quantum confinement of the
  conduction electrons (superlattice modulation) leads to a complex, quasi-two dimensional subband
  structure with both hole- and electron-like Fermi surfaces. ii) strong
  electron-electron interaction requires a substantial renormalization of
  the quasi-particle dispersion. We characterize this renormalization by two sets of parameters, namely, the quasi-particle weight and the induced particle-hole asymmetry of each partially filled subband. 
In our study, the quasi-particle dispersion is calculated self-consistently as function of microscopic parameters using the slave-boson mean-field approximation introduced by Kotliar and Ruckenstein. 
We discuss the consequences of strong local correlations on the normal-state free-carrier response in the optical conductivity and on the thermoelectric effects.}

\section{Introduction}
\label{sec:intro}
Recent experiments \cite{Ohtomo:2002fk} have shown that a metallic state can be stabilized at the interface between the Mott insulator LaTiO$_3$ and the band insulator SrTiO$_3$. There is strong evidence that in such systems electronic charge is redistributed between the Mott insulator (MI) and the band insulator (BI) in order to compensate for the mismatch of the work functions and to avoid the so-called polar catastrophe \cite{nakagawa:2006}. The electronic charge reconstruction \cite{Okamoto:2004uq} at the interface leads to metallic behavior.

\begin{figure}
\sidecaption[t]
\includegraphics[width=0.6\linewidth]{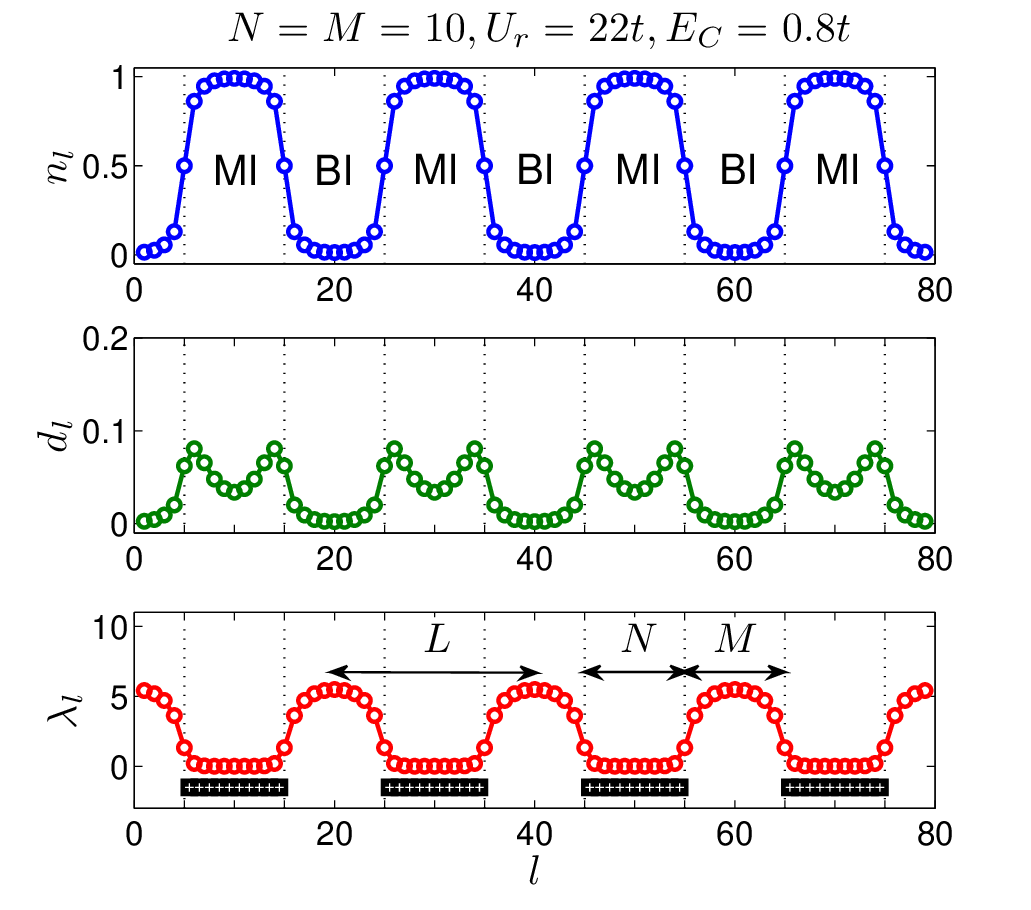}
\caption{The charge density $n_l$, the fraction of doubly occupied sites $d_l^2$ and the Lagrange multiplier $\lambda_l$ as obtained by the present mean-field approach for a band-insulator (BI)/Mott-insulator (MI) superlattice. $L=N+M$ denotes the number of unit cells of the superlattice modulation where $N$ is the number of MI-layers and $M$ the number of BI-layers. The set of microscopic parameters is given in the figure.}
\label{fig:sol}
\end{figure}

The experimental data \cite{Ohtomo:2002fk, Seo:2007, takizawa:057601,Shibuya:2004} are consistent with Fermi liquid behavior and a single-particle perspective, where transport properties are studied by the semiclassical transport equations, offers a natural starting point. However, it is necessary to clarify how the single-particle picture for weakly interacting electrons, which is successfully applied in the study of semiconductor nano-structures, is modified by strong electronic correlations. In particular, understanding the renormalization of the quasi-particle dispersion as function of microscopic parameters is crucial. Let us in the following ignore complicating aspects related to the orbital degrees of freedom \cite{J.Chakhalian11162007} or to possible symmetry-broken phases \cite{Okamoto:2004uq}. Then, from quite general considerations, we can expect two major ingredients determining the electronic structure: 

(i) The LaTiO$_3$ bulk system has a Ti-3$d^1$ configuration whereas the SrTiO$_3$ compound has a Ti-3$d^0$ configuration. Therefore, from a single-particle point of view, the BI/MI/BI sandwich acts as a quantum well confining the conduction electrons to the MI region, cf. Fig.~\ref{fig:sol}. The bound states of the quantum well form \emph{quasi-two dimensional subbands} with dispersion $E_{{\bf k}\nu}$ labeled by the in-plane momentum $\bf k$ and the subband index $\nu$.\footnote{Our discussion is restricted to large superlattice periods where the dependence on the perpendicular momentum $Q$ can be neglected when studying the parallel transport.} The Fermi surface (FS) defined by the ${\bf k}$ points satisfying $E_{{\bf k}\nu}=0$ contains in general both open and closed sheets which brings about that electron-like and hole-like contributions can lead to partial compensation \cite{Pippard:1989}.

(ii) \emph{Strong electron-electron interaction} introduces novel electronic physics at the band-insulator/ Mott-insulator interface. When the local self-energy corrections are dominant, we can assume that $E_{{\bf k}\nu}=E_{\nu}(\varepsilon_{\bf k})$, where $\varepsilon_{\bf k}$ is the non-interacting in-plane dispersion \cite{rueegg:2007, rueegg:2008}. In this case, the renormalization of the quasi-particle dispersion is characterized by two sets of parameters. On the one hand, the on-site repulsion leads to a reduction of the Fermi velocity of the subband $\nu$ by a factor
\begin{equation}
Z_{\nu}=\left.\frac{\partial E_{{\bf k}\nu}}{\partial\varepsilon_{\bf k}}\right|_{\rm FS}
\label{eq:Z}
\end{equation}
which is equal to the quasi-particle weight of the subband $\nu$. On the other hand, at the interface, the hybridization of (almost) localized with itinerant degrees of freedom induces an enhanced particle-hole asymmetry. We quantify this asymmetry by the dimensionless parameter
\begin{equation}
\alpha_{\nu} = \left.\left(\Delta\varepsilon_{\bf k}\frac{\partial^2 E_{{\bf k}\nu}}{\partial \varepsilon_{\bf k}^2}\right)/\left(\frac{\partial E_{{\bf k}\nu}}{\partial \varepsilon_{\bf k}}\right)\right|_{{\rm FS}},
\label{eq:alpha}
\end{equation}
where we have defined $\Delta\varepsilon_{\bf k}=\varepsilon_{\bf k}-\varepsilon_{b}$ with $\varepsilon_b$ the energy of the lower band edge.
Starting from a microscopic model, we calculate the quasi-particle dispersion in a self-consistent way to obtain the dependence on microscopic parameters. We obtain an \emph{interfacial heavy-fermion state} and discuss correlation effects as characterized by Eq.~(\ref{eq:Z}) and (\ref{eq:alpha}), cf. Fig.~\ref{fig:EnZa}. Implications for transport are illustrated by calculating the free carrier response and the thermoelectric effects, cf. Fig.~\ref{fig:DQ}. We find that in both cases important contributions arise from the interface.

\section{Microscopic model}
\label{sec:model}
For the microscopic description we assume perfect lattice match between the two materials, thereby neglecting aspects related to the lattice relaxation \cite{okamoto:056802}. The microscopic model is given by an extended single-orbital Hubbard model on a cubic lattice (introduced in Ref.~\cite{okamoto:241104}) 
\begin{equation}
H=H_{t}+H_{U}+H_{ee}+H_{ei}+H_{ii}.
\label{eq:H}
\end{equation}
Here, the kinetic energy is given by a nearest-neighbor tight-binding model and the on-site repulsion is modeled by a Hubbard interaction,
\begin{equation}
H_{t}=-t\sum_{\langle ij\rangle,\sigma}c_{i\sigma}^{\dag}c_{j\sigma}^{}+{\rm h. c.}\quad\mathrm{and} \quad H_{U}=U\sum_i n_{i\uparrow}n_{i\downarrow},
\end{equation}
where $n_{i\sigma}=c_{i\sigma}^{\dag}c_{i\sigma}^{}$. The nanoscale structure is defined by the superlattice period $L=N+M$ and the number $N$ of counter-ion layers (see also Fig.~\ref{fig:sol}). They simulate the difference between Sr$^{2+}$ and La$^{3+}$ and sit in the center between the electronic sites \cite{Okamoto:2004uq} interacting with the electrons through the long-range electron-ion interaction
\begin{equation}
H_{ei}=-E_C\sum_{i,j}\frac{n_i}{|\vec{r}_i-\vec{r}_j^{\rm ion}|}
\end{equation}
where $\vec{r}_j^{\rm ion}$ denotes the position of the ions, $n_i=n_{i\uparrow}+n_{i\downarrow}$, and we have introduced the parameter $E_C$ controlling the screening length. Furthermore, the long-range electron-electron and ion-ion interaction energies are given by
 \begin{equation}
H_{ee}=\frac{E_C}{2}\sum_{i\neq j}\frac{n_in_j}{|\vec{r}_i-\vec{r}_j|},\quad H_{ii}=\frac{E_C}{2}\sum_{i\neq j}\frac{1}{|\vec{r}_i^{\rm ion}-\vec{r}_j^{\rm ion}|},
 \end{equation} 
respectively. The number of electrons is fixed by the charge-neutrality condition. Notice that we can formally relate the parameter $E_C$ to an effective dielectric constant $\epsilon=e^2/E_Ca$ where $a$ is the lattice constant and $e>0$ the elementary charge. However, for a more realistic description of the screening at the interface a single parameter for the long-range electron-electron interaction is too crude. In fact, the polarization of the lattice dominates the dielectric constant in the considered transition metal oxides and the effect of the relaxation of the lattice near the interface introduces additional parameters in an effective model description \cite{okamoto:056802, Hamann:2006}. For simplicity, such effects are not considered here.

\section{Slave-boson mean field approximation}
\label{sec:effmod}
To discuss the low-energy behavior of the model (\ref{eq:H}) in the normal state we apply
the four-boson mean-field approximation of Kotliar and Ruckenstein \cite{Kotliar:1986kx}. This approach allows to discuss the effect of local self-energy corrections by introducing auxiliary bosons representing the local charge and spin degrees of freedom together with pseudo fermions. The effective low-energy theory is then obtained by the saddle-point approximation for the slave bosons which can be controlled by a $1/N$-expansion of a suitable generalization of the slave-boson action \cite{Fresard:1992a, arrigoni:1995}. The remaining fermionic degrees of freedom are interpreted as the Landau quasi-particles of a Fermi liquid which are dressed by the interactions and therefore have modified single-particle properties.

\subsection{Superlattice geometry}
\label{subsec:sup}
For a quantum well system, the effective low-energy model was derived in Ref.~\cite{rueegg:2007}. In the following, we will briefly discuss the case of a superlattice. Assuming a translational invariant state in the in-plane direction, the problem of finding the eigenvalues of the effective low-energy Hamiltonian reduces to a one-dimensional problem which is parameterized by the non-interacting dispersion $\varepsilon_{\bf k}=-2t(\cos k_x+\cos k_y)$:
\begin{equation}
(z_l^2\varepsilon_{\bf k}+\lambda_l)\psi(l)-t\sum_{\gamma=\pm 1}z_lz_{l+\gamma}\psi(l+\gamma)=E\psi(l).
\label{eq:SE}
\end{equation}
Here, $l$ labels the layers along the direction of the superlattice modulation. The hopping renormalization amplitude $z_l$ depends on the charge density $n_l$ and the fraction of doubly occupied sites $d_l^2$ in layer $l$ and is given by the standard Gutzwiller expression \cite{Gutzwiller:1963lr}
\begin{equation}
z_l=\frac{\sqrt{(1-n_l+d_l^2)(n_l-2d_l^2)}+d_l\sqrt{n_l-2d_l^2}}{\sqrt{n_l(1-n_l/2)}}.
\end{equation}
The mean fields $n_l$ and $d_l$ are determined by the minimum of an appropriate free energy, as discussed below. The Lagrange multiplier $\lambda_l$ acts as a single-particle potential and enforces the self-consistency of the electronic charge distribution.

Decomposing $\psi$ according to $\psi(l)=\psi_{\vec{K}\nu}(l)e^{iQl}$, where $\vec{K}=({\bf k},Q)$, $-\pi/La\leq Q<\pi/La$ with $\psi_{\vec{K}\nu}(l+L)=\psi_{\vec{K}\nu}(l)$, the problem reduces to diagonalizing the following matrix
\begin{equation}
\hat{K}(Q)=\left(\begin{array}{cccc}
z_1^2\varepsilon_{\bf k}+\lambda_1&-tz_1z_2e^{iQ} &\dots&-tz_1z_Le^{-iQ}\\ 
-tz_2z_1e^{-iQ}&z_2^2\varepsilon_{\bf k}+\lambda_2 &-tz_2z_3e^{iQ}&\dots\\
\dots & \dots & \dots &\dots \\
-tz_Lz_1e^{iQ} &\dots&-tz_{L-1}z_Le^{-iQ}& z_L^2\varepsilon_{\bf k}+\lambda_L
\end{array}\right).
\end{equation} 
A further simplification is obtained by restricting to superlattices with a large period $L\gg 1$. In this case, the $Q$-dependence can be safely neglected for the parallel transport and it is sufficient to consider only $\hat{K}(0)$. In this case, the quasi-particle dispersion has the form stated in the introduction, $E=E_{\nu}(\varepsilon_{\bf k})$, $\nu=1,\dots,L$. 

\subsubsection{Long range Coulomb interaction}
The long-range Coulomb interaction is treated in the Hartree-type of mean-field calculation \cite{rueegg:2007}. In order to find the interaction matrix $W_{ll'}$ between electrons in layer $l$ and $l'$ of the superlattice unit cell we explicitly take into account the periodicity of the charge distribution. In the same way, we also determine the resulting (screened) potential $V_l$ of the counter-ions. In the end, these potentials have to be found self-consistently by simultaneously solving the mean-field equations and the Poisson equation, $\Delta\phi(z)=-4\pi\frac{\rho(z)}{\epsilon}$, where $\phi(z)$ is the electrostatic potential and $\rho(z)$ the charge distribution. 

We start with the approximation commonly found in the literature \cite{Freericks:2006}, namely, we replace each layer by a uniformly charged plane, thereby respecting the polar nature of the Mott insulator. Corrections due to the discrete nature of the charge distribution (lattice) are calculated numerically, but are only significant very close to (or within) the considered layer \cite{wehrli:2001}. It is convenient to use the following elementary solution $\phi_o$ determined by a periodical array of uniformly charged layers with period $La$
\begin{equation}
\rho_o(z)=\sigma\sum_{m\in Z}\delta(z+mLa)-\bar{\rho}
\end{equation}
where $\sigma=e/a^2$ is an elementary surface-charge density and $\bar{\rho}=\sigma/La$ is a uniform background charge to keep the total system charge neutral. The solution can be written in a compact form by use of the polylogarithm ${\rm Li}_b(z)=\sum_{n>0}\frac{z^n}{n^b}$:
\begin{equation}
\label{eq:elsol}
\phi_o(z)=\frac{\sigma La}{\epsilon\pi}\left[{\rm Li}_2(e^{2\pi i z/La})+{\rm Li}_2(e^{-2\pi i z/La})\right].
\end{equation}
Between two neighboring layers at $nLa$ and $(n+1)La$, the resulting potential is simply given by the parabola
\begin{equation}
\phi_o(z)=\frac{2\pi\sigma}{L\epsilon}(z-nLa)(nLa+La-z).
\end{equation}
From the elementary solution Eq.~(\ref{eq:elsol}) it is straightforward to determine $W_{ll'}$ and $V_l$ by summing up the contributions from the different layers within the superlattice unit cell and adding the numerically determined correction terms due to the discreteness of the charge distribution.

\subsubsection{Free energy}
Eventually, we give the expression for the free energy per lattice site ($\beta^{-1}=k_BT$)
\begin{equation}
f(n,d,\lambda)=\frac{-2}{\beta N_{||}L}\sum_{{\bf k}\nu}\ln(1+e^{-\beta E_{{\bf k}\nu}})+\frac{U_r}{L}\sum_l{d_l}^2+\frac{1}{2L}\sum_{ll'}n_lW_{ll'}n_{l'}-\frac{1}{L}\sum_l(\lambda_l-V_l)n_l.
\label{eq:f}
\end{equation} 
Sums over layers are restricted to a single superlattice unit cell. The self-consistency equations are solved by maximizing Eq.~(\ref{eq:f}) with respect to $\lambda_l$ and minimizing the resulting function with respect to the mean fields $n_l$ and $d_l$ under the constraint of charge neutrality, $\sum_ln_l=N$. 

A typical solution of the self-consistency equations at $T=0$ is shown in Fig.~\ref{fig:sol} for a superlattice with $N=M=10$. The charge distribution $n_l$ allows to naturally distinguish between three different regions, $n_l\approx 1$, $n_l\approx 0.5$ and $n_l\approx 0$. The layers with filling $n_l\approx0.5$ seperate the ``Mott-insulating" (MI) regions ($n_l\approx1$) from the ``band-insulating" (BI) regions with $n_l\approx 0$ (notice that the whole system is actually metallic). The presence of the interface (IF) layers with filling $n_l\approx0.5$ is a consequence of the polar nature of the Mott insulator.
\begin{figure}
\sidecaption[t]
\includegraphics[width=0.6\linewidth]{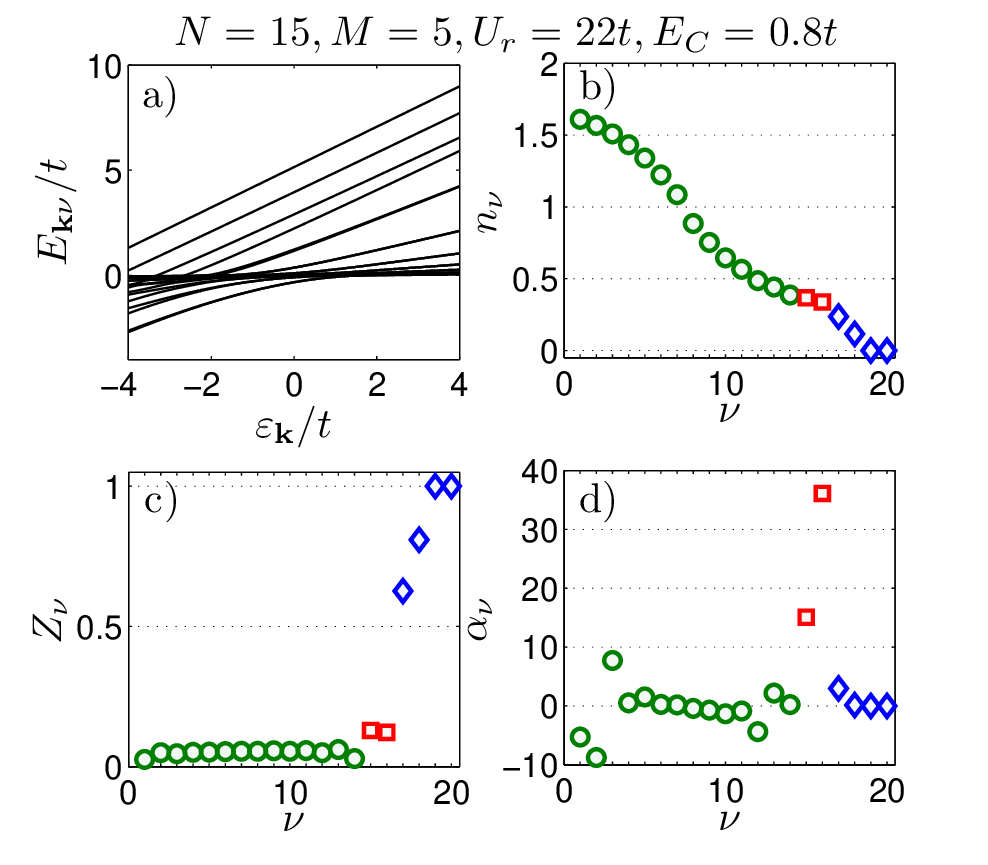}
\caption{a) The quasi-particle dispersion $E_{{\bf k}\nu}$ as function of the non-interacting in-plane dispersion $\varepsilon_{\bf k}$. b) The subband filling $n_{\nu}$, c) the quasi-particle weight $Z_{\nu}$ and d) the induced particle-hole asymmetry $\alpha_{\nu}$ for the individual subbands which correspond to the dispersion in a). Circles (green) are associated with the subbands of the MI region, squares (red) with the IF and diamonds (blue) with the BI region.}
\label{fig:EnZa}
\end{figure}
\section{Transport properties}
From the self-consistent solution of the mean-field equations we also obtain the quasi-particle dispersion $E_{{\bf k}\nu}$ and the envelope wave-function $\psi_{{\bf k}\nu}(l)$. At low temperatures, it is expected that the transport properties can be understood from the properties of the quasi-particles \cite{rueegg:2008}. We therefore start by discussing the electronic structure.  
\subsection{Characterization of the generic electronic structure}
The quasi-particle dispersion $E_{{\bf k}\nu}$ is shown in Fig.~\ref{fig:EnZa}~a) for the $N=15$, $M=5$ superlattice. The value of the on-site repulsion, $U_r=22t$, is well above the critical interaction strength of the Mott transition in the half-filled bulk system \cite{Brinkman:1970lr}, $U_c\approx 16t$. Thus, the quasi-particle dispersion shown in Fig.~\ref{fig:EnZa} corresponds to the strongly correlated regime. In panels b) to d) we show the subband filling $n_{\nu}$, the quasi-particle weight $Z_{\nu}$ [Eq.~(\ref{eq:Z})] and the induced particle-hole asymmetry $\alpha_{\nu}$ [Eq.~(\ref{eq:alpha})] of the subbands shown in a). Although the spatial weight of the envelope wave-function $\psi_{{\bf k}\nu}(l)$ extends over the whole super unit cell and also depends on the value of $\varepsilon_{\bf k}$ \cite{rueegg:2007}, it is possible to group the different subband states according to the regions where most of their spatial weight is located. We thus define $\nu_{\rm MI}=1,\dots,N-1$, $\nu_{\rm IF}=N,N+1$ and $\nu_{\rm BI}=N+2,\dots,L$. Notice that due to strong local correlations, the quasi-particle weight $Z_{\nu_{\rm MI}}$ and $Z_{\nu_{\rm IF}}$ is strongly reduced for the subbands of the MI and IF region [panel c)], whereas the particle-hole asymmetry $\alpha_{\nu_{\rm IF}}$ is enhanced most dominantly for the subbands of the IF region [panel d)]. These are the basic characteristics of the \emph{interfacial heavy-fermion state} obtained by the slave-boson mean-field approximation at low temperatures. The coherent hybridization of the itinerant degrees of freedom in the BI and IF region with the almost localized degrees of freedom in the MI region is mediated by the intra-layer hopping and leads to heavy-fermion behavior of the interfacial subbands. The situation is  reminiscent of heavy-fermion systems as described for example by the periodic Anderson model \cite{Rice:1985lr}. However, in the present case, localized and itinerant degrees of freedom have the same orbital character but are separated spatially - in contrast to the classical heavy-fermion systems where the localized f-electrons hybridize with the states of the conduction band.  
\begin{figure}
\centering
\includegraphics[width=0.49\linewidth]{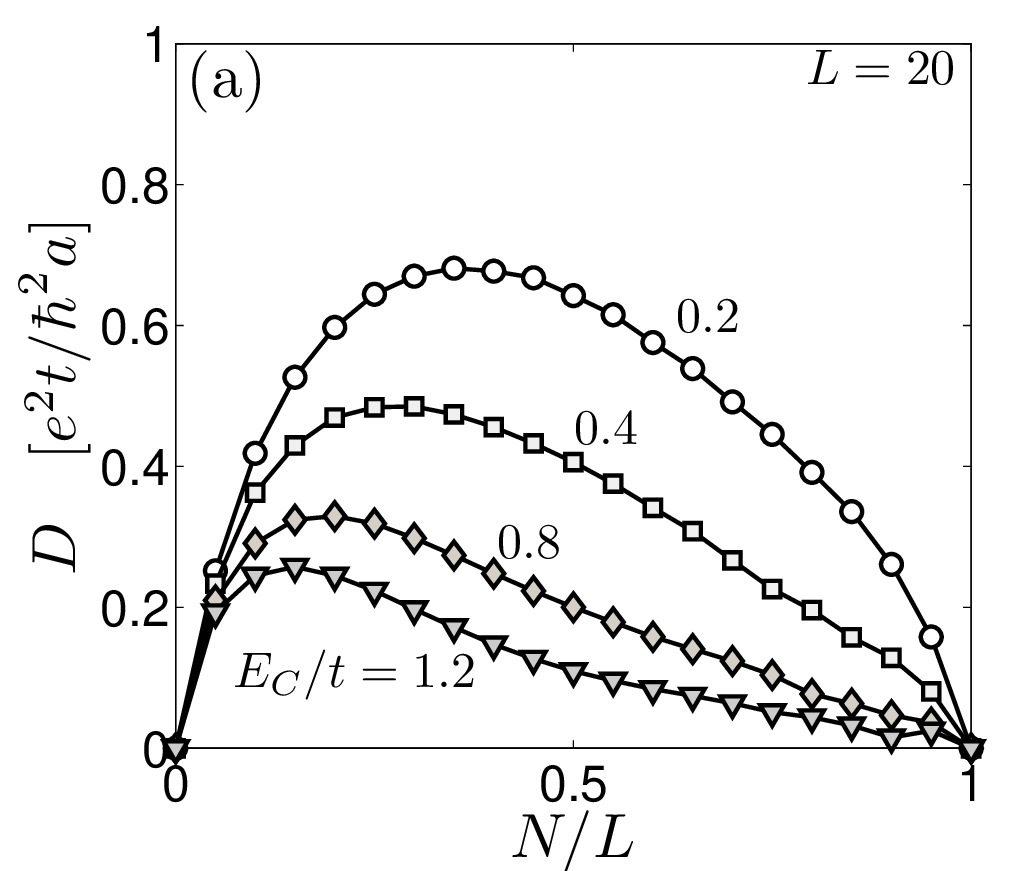}
\includegraphics[width=0.49\linewidth]{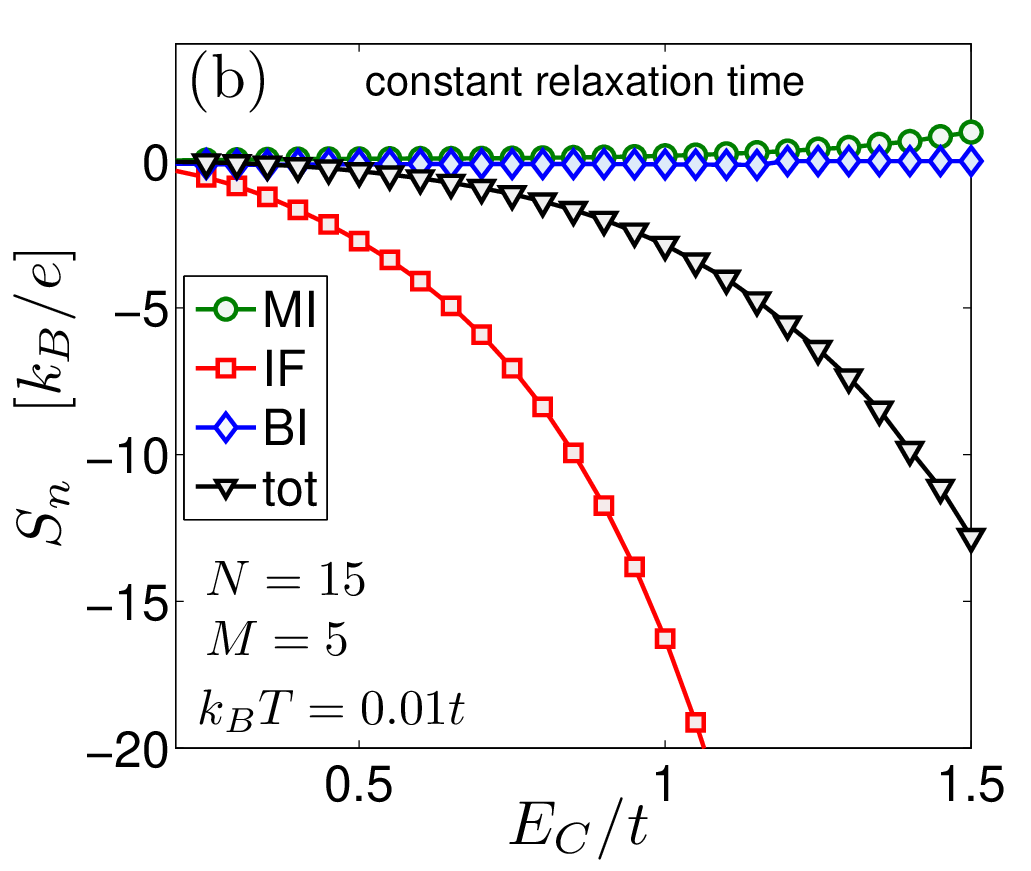}
\caption{(color online). (a) The Drude weight $D$ as function of $N/L$ for different values of $E_C$ for a superlattice with period $L=20$. (b) The total Seebeck coefficient $S$ (triangles, black) and the different contributions $S_n$ associated with the MI (circles, green), IF (squares, red) and BI (diamonds, blue) regions as a function of $E_C$ evaluated for a constant relaxation time. In both panels, the value of the on-site interaction is fixed at $U_r=22t$.}
\label{fig:DQ}
\end{figure}
\subsection{Drude weight}
The \emph{Drude weight} in the optical conductivity is obtained as described in \cite{rueegg:2007, rueegg:2008}. One finds the familiar expression for a quasi-particle response,
\begin{equation}
D=\sum_{\nu}D_{\nu},\quad D_{\nu}=\frac{e^2}{4\pi L a}A_{\nu}\bar{v}_{\nu},
\end{equation}
where $A_{\nu}$ is the Fermi surface volume of the sheet $\nu$ and $\bar{v}_{\nu}=Z_{\nu}\langle|\nabla_{\bf k}\varepsilon_{\bf k}|\rangle_{\rm FS}/\hbar$ is the Fermi velocity averaged over the Fermi surface. In Fig.~\ref{fig:DQ}~a) we illustrate how $D$ evolves from the band insulator ($N=0$) to the Mott insulator ($N=L$), which both have a vanishing $D$. The maximal $D$ as a function of the averaged electronic density $N/L$ depends on the value of $E_C/t$ and shifts to lower $N$'s for increasing $E_C$ because the screening length is reduced. 

\subsection{Seebeck coefficient}
Thermoelectric effects are characterized by the Seebeck coefficient $S$. In terms of the subband contributions it is written as
\begin{equation}
S=\sum_{\nu}\frac{S_{\nu}\sigma_{\nu}}{\sigma},\quad\sigma=\sum_{\nu}\sigma_{\nu}.
\end{equation}  
Here, $\sigma$ is the total electrical conductivity and $S_{\nu}$ the Seebeck coefficient associated with the subband $\nu$.
In lowest order in the temperature we have \cite{rueegg:2008}
\begin{equation}
S_{\nu}=-\frac{\pi^2}{3}\frac{k_B}{e}\frac{k_BT}{Z_{\nu}\Delta\varepsilon_{\nu}^*}\left[\alpha_{\nu}+\Delta\varepsilon_{\nu}^*\left(\frac{\tau_{\nu}'}{\tau_{\nu}}+\frac{\mathcal{N}_v'}{\mathcal{N}_v}\right)\right]
\label{eq:Qnu}
\end{equation}
where $\Delta\varepsilon_{\nu}^*=\varepsilon_{\nu}^*-\varepsilon_b$ is measured from the band edge and defined through $E_{\nu}(\varepsilon_{\nu}^*)=0$, $\tau_{\nu}(\varepsilon)$ is the relaxation time of the subband $\nu$ and 
\begin{equation}
\mathcal{N}_v(\varepsilon)=\int\frac{d^2k}{(2\pi)^2}|\nabla_{\bf k}\varepsilon_{\bf k}|^2\delta(\varepsilon-\varepsilon_{\bf k}).
\end{equation}
The prime ($'$) in Eq.~(\ref{eq:Qnu}) denotes the derivative with respect to $\varepsilon$ at the Fermi surface. Attempts to calculate $\tau_{\nu}$ from a microscopic model offers a challenging task in correlated, disordered and inhomogeneous systems (see also Ref.~\cite{rueegg:2008}). For simplicity, we discuss here the case of a constant relaxation time. We assume an energy-independent relaxation time $\tau_n$ for the subbands associated with the different regions, $n=$ MI, IF, BI, and therefore obtain
\begin{equation}
S=\sum_n\frac{S_n\sigma_n}{\sigma},\quad S_n=\sum_{\nu_n}\frac{S_{\nu}D_{\nu}}{D_n},\quad D_n=\sum_{\nu_n}D_{\nu},\quad \sigma_n=D_n\tau_n.
\end{equation}
Figure~\ref{fig:DQ}~b) shows the different contributions $S_n$ to the total Seebeck coefficient for a superlattice with $N=15$ and $M=5$ as a function of the parameter $E_C$. The contribution $S_{\rm IF}$ from the subbands associated with the interface is most dominant and $|S_{\rm IF}|$ increases for increasing $E_C$ (remember that a large value of $E_C$ yields a sharp charge distribution). This can be understood by the fact that the particle-hole asymmetry $\alpha_{\nu}$ induced by the correlations is largest for the subbands of the interface region and increases for a sharper interface due to a reduction in the hybridization (intra-layer hopping). However, the total (absolute) Seebeck coefficient is at best equal to its largest subband contribution, $|S|\leq\max_{\nu}|S_{\nu}|$, but is in general smaller, as shown in Fig.~\ref{fig:DQ}~b) by assuming $\tau_{\rm MI}=\tau_{\rm IF}=\tau_{\rm BI}$. Nevertheless, we can state the conditions for which the interface contribution becomes large in the constant-relaxation-time approximation: (i) strong electronic correlations, $U_r>U_c$. (ii) large values of $N$, such that bulk-like properties in the center of the MI are obtained. (iii) a sharp interface ($E_C>t$).
\subsection{Comparison to the atomic limit result}
Note that the above discussed enhancement of $|S_{\rm IF}|$ at low temperatures is a non-local effect. It is therefore expected that any reduction of spatial coherence across the interface will suppress and eventually destroy this mechanism. The influence of the reduction of spatial coherence due to thermal fluctuations is clearly seen by analyzing $S$ in the high-temperature (atomic) limit. Let us in the following discuss the situation $t\ll k_BT\ll U$ such that doubly occupied sites are completely suppressed. In the {\em homogeneous} system at density $n$, the thermopower $S$ is given by the entropic contribution alone \cite{Beni:1974, Chaikin:1976}. This yields Heikes formula
\begin{equation}
\label{eq:Heikes}
S=-\frac{k_{\rm B}}{e}\log\left[\frac{2(1-n)}{n}\right]
\end{equation}
where the $\log 2$ contribution arises from the entropy of the spin degree of freedom. As pointed out in Ref.~\cite{Koshibae:2000}, in transition metal oxides also the inclusion of orbital degrees of freedom is necessary. This is in principle straight forward but we will not discuss it further here. For the {\em inhomogeneous} system an appropriate generalization of Eq.~(\ref{eq:Heikes}) is given by
\begin{equation}
\label{eq:Shigh}
S=\frac{\sum_lS_l\sigma_l}{\sigma}=-\frac{k_B}{e}\frac{\sum_l\log\left[\frac{2(1-n_l)}{n_l}\right]n_l(1-n_l)}{\sum_ln_l(1-n_l)}.
\end{equation}
Here we have used the {\em local} quantities
\begin{equation}
S_l=-\frac{k_{\rm B}}{e}\log\left[\frac{2(1-n_l)}{n_l}\right],\quad \sigma_l=\frac{e^2 A\beta}{2}n_l(1-n_l),
\end{equation}
with a temperature and doping independent constant $A$ \cite{Mukerjee:2007}. The weighted sum in Eq.~(\ref{eq:Shigh}) clearly shows that an inhomogeneous system is not favorable as long as a {\em local} description is appropriate. In fact, if one seeks to optimize the powerfactor PF$=S^2\sigma$ in this limit for a spatially varying density profile, the optimal solution is found to be the homogeneous solution with optimal density $n\approx 0.12$, c.f. Ref.~\cite{Mukerjee:2007}. This is exactly the opposite behavior than found in the low-temperaure limit. Restricting to purely electronic contribution, we therefore conclude that a spatially non-uniform system can only be favorable if spatial coherence is sustained.

\section{Conclusions}
In summary, we have studied aspects of the parallel transport at low-temperatures in strongly-correlated superlattices from the semiclassical point of view. The generic electronic structure is discussed and self-consistently computed from microscopic parameters using the four-boson approach of Kotilar and Ruckenstein to deal with strong local correlations. Implications for the parallel transport are illustrated by the free-carrier response and the thermoelectric effects. The presence of the interface introduces new aspects not feasible in the bulk systems. Here, we have discussed the scenario of an interfacial heavy-fermion state where the coherent hybridization of itinerant and almost localized degrees of freedom leads to a large particle-hole asymmetry which can be responsible for a high value of the Seebeck coefficient. We find that this mechanism is a non-local effect and that spatial coherence is crucial. If such mechanisms are also relevant for the giant thermopower observed in SrTiO$_3$/SrTi$_{0.8}$Nb$_{0.2}$O$_3$ superlattices \cite{Ohta:2007} need to be clarified by further studies of more realistic models.

\begin{acknowledgement}
We would like to thank S.~Pilgram, M.~Ossadnik, R. Asahi and T.~M.~Rice for valuable discussions.
We acknowledge financial support of the Toyota Central R\&D Laboratories, Nagakute, Japan and the NCCR MaNEP of the Swiss Nationalfonds.
\end{acknowledgement}


\begin{thebibliography}{99.}%

\bibitem{Ohtomo:2002fk}
A.\ Ohtomo, D.A.\ Muller, J.L.\ Grazul, and H.Y.\ Hwang,
 Nature {\bf 419}, 378 (2002).

\bibitem{nakagawa:2006}
N.\ Nakagawa, H.Y.\ Hwang, and D.A.\ Muller,
 Nature Mat.\ {\bf 5}, 204 (2006).

\bibitem{Okamoto:2004uq}
S.\ Okamoto and A.J.\ Millis,
 Nature {\bf 428}, 630 (2004).

\bibitem{Seo:2007}
S.S.A.\ Seo {\em et~al.},
 Phys.\ Rev.\ Lett.\ {\bf 99}, 266801 (2007).

\bibitem{takizawa:057601}
M.\ Takizawa {\em et~al.},
 Phys.\ Rev.\ Lett.\ {\bf 97}, 057601 (2006).

\bibitem{Shibuya:2004}
K.\ Shibuya {\em et~al.},
 Jpn.\ J.\ Appl.\ Phys.\ {\bf 43}, L1178 (2004).

\bibitem{J.Chakhalian11162007}
J.\ Chakhalian {\em et~al.},
 Science {\bf 318}, 1114 (2007).

\bibitem{rueegg:2007}
A.\ R\"uegg, S.\ Pilgram, and M.\ Sigrist,
 Phys.\ Rev.\ B {\bf 75}, 195117 (2007).

\bibitem{rueegg:2008}
A.\ R\"uegg, S.\ Pilgram, and M.\ Sigrist,
 Phy.\ Rev.\ B {\bf 77}, 245118 (2008).

 \bibitem{Pippard:1989}
 A.\ B.\ Pippard
{\em Magnetotransport in metals}
(Cambridge University Press, 1989)

\bibitem{okamoto:056802}
S.\ Okamoto, A.J.\ Millis, and N.A.\ Spaldin,
 Phys.\ Rev.\ Lett.\ {\bf 97}, 056802 (2006).
 
\bibitem{Hamann:2006}
D.\ R.\ Hamann, D.\ A.\ Muller, and H.\ Y.\ Hwang
Phys.\ Rev.\ B {\bf 73}, 195403 (2006)

\bibitem{okamoto:241104}
S.\ Okamoto and A.J.\ Millis,
 Phys.\ Rev.\ B {\bf 70}, 241104(R) (2004).

\bibitem{Kotliar:1986kx}
G.\ Kotliar and A.E.\ Ruckenstein,
 Phys.\ Rev.\ Lett.\ {\bf 57}, 1362 (1986).

\bibitem{Fresard:1992a}
R.\ Fr\'esard and P.\ W\"olfle,
 Int.\ J.\ Mod.\ Phys.\ B {\bf 6}, 237 (1992).

\bibitem{arrigoni:1995}
E.\ Arrigoni and G.C.\ Strinati,
 Phy.\ Rev.\ B {\bf 52} (1995).

\bibitem{Gutzwiller:1963lr}
M.C.\ Gutzwiller,
 Phys.\ Rev.\ Lett.\ {\bf 10}, 159 (1963).

\bibitem{Freericks:2006}
J.K.\ Freericks,
 {\em Transport in multilayered nanostructures: the dynamical mean-field
  approach}
 (Imperial College Press, London, 2006).

\bibitem{wehrli:2001}
S.\ Wehrli, D.\ Poilblanc, and T.M.\ Rice,
Eur.\ Phys.\ J.\ B {\bf 23}, 345 (2001).

\bibitem{Brinkman:1970lr}
W.F.\ Brinkman and T.M.\ Rice,
 Phys.\ Rev.\ B {\bf 2}, 4302 (1970).

\bibitem{Rice:1985lr}
T.M.\ Rice and K.\ Ueda,
 Phys.\ Rev.\ Lett.\ {\bf 55}, 995 (1985).
 
\bibitem{Beni:1974}
G.\ Beni
Phys. Rev. B {\bf 10}, 2186 (1974)

\bibitem{Chaikin:1976}
P.\ M.\ Chaikin and G.\ Beni
Phys. Rev. B {\bf 13}, 647 (1976)

\bibitem{Koshibae:2000}
W.\ Koshibae, K.\ Tsutsui, and S.\ Maekawa 
Phys. Rev. B {\bf 62}, 6869 (2000)

\bibitem{Mukerjee:2007}
S.\ Mukerjee and J.\ E.\ Moore
Appl. Phys. Lett {\bf 90}, 112107 (2007)

\bibitem{Ohta:2007}
H.\ Ohta {\em et~al.},
Nat. Mat. {\bf 6}, 129 (2007)

\end{thebibliography}
\end{document}